# ANONYMER TANZ ALS DEKOLONIALISIERENDE PRAXIS. EIN EMBODIED-RESEARCH VERSUCH

Paula Helm

Dieser Aufsatz widmet sich der Frage, wie und mit welchen Folgen Anonymität als ein gestalterisches Element in Experimenten eingesetzt wird, bei denen es um alternative Formen der Wissensproduktion geht. Spezifischer richtet sich der Blick auf eine bestimmte Strömung im zeitgenössischen Tanz, den experimentellen Tanz. Diese hat sich im Laufe der 1970er Jahre entwickelt.[1] Ihre Leitidee ist es, den Fokus von der fertigen Choreographie und Aufführung weg auf die Prozesshaftigkeit, Spontanität und Erfahrung der tänzerischen Praxis selbst zu rücken.[2] Tanz wird dabei als ein Laboratorium verstanden, in dem mit Schwerkraft, Körpergrenzen, Gruppendynamik und Wahrnehmung experimentiert wird. Im Zentrum steht das Testen, Tasten, Probieren.[3] Die entsprechenden Veranstaltungen nennen sich je nach Regelmäßigkeit entweder ›Dance Labs‹ oder ›Movement Research Workshops‹.

Ein Instrument, das im Rahmen derartiger Veranstaltungen häufig verwendet wird, um spezifische soziale Situationen und Gruppendynamiken zu erzeugen oder um Wahrnehmungen in bestimmten Richtungen zu kanalisieren, ist die Anonymität. Anonymität wird dabei weit gefasst als das Kappen von Verbindungen verstanden, welches eine gewisse Unzurechenbarkeit in Bezug auf unterschiedliche Aspekte einer Person, einer Situation oder einer Tätigkeit zur Folge hat. Beispielsweise kann Unzurechenbarkeit dadurch entstehen, dass Personen sich gegenseitig nicht mehr sehen können, so dass optische Identifikation, wenn teils auch nur temporär, unmöglich wird. Durch das Durchbrechen gewohnter Modi der Identifikation, können zuweilen andere Formen des In-Beziehung-Tretens hervorgehoben werden. Das macht diese Form der Anonymität so interessant für den experimentellen Tanz.

Auf Basis eines derart breiten Verständnisses von Anonymität, diskutiert dieser Beitrag einige Einblicke aus einer Studie, in deren Rahmen ich mich zum einen mit der Frage beschäftigt habe, ›wie‹ Anonymität in wissenschaftlichen Tanz-Experimenten eingesetzt und gestaltet wird und zum anderen mit der Frage, ›welche‹ Motive, Imaginäre und theoretischen Hintergrundüberlegungen diesen Experimenten zu Grunde liegen. Dazu beginne ich zunächst mit einer kurzen Beschreibung des For-schungsgegenstandes sowie mit einer Methodenreflexion, bevor ich drei Fallbeispiele einer eingehenderen Analyse unterziehe.

---

1 *Cynthia J. Novack:* Sharing the Dance. Contact Improvisation and American Culture. Wisconsin 1990.
2 *Ann Cooper Albright/David Gere* (Hg.): Taken by Surprise. A Dance Improvisation Reader. Middletown 2003.
3 *Gabriele Klein:* Künstlerische Praktiken des Ver(un)sicherns. Produktionsprozesse am Beispiel des Tanztheater Wuppertal Pina Bausch. In: Paragrana 25 (2015), S. 201–208, hier S. 206.

## Tanz als alternativer Modus der Wissensproduktion

Über Tanz und insbesondere über experimentellen Tanz zu sprechen und zu schreiben ist mit methodologischen Herausforderungen verbunden. Der Grund dafür ist, dass es eine der basalen Ideen des experimentellen Tanzes ist, Formen des Ausdrucks, der Kommunikation und der Wissensproduktion zu mobilisieren, die nicht den Limitationen des sprachlichen Ausdrucks unterliegen. Die Majorität sogenannter »standard modes of knowing«[4] basiert auf Text und Sprache als Medien der Wissensproduktion und des Wissenstransfers. Das gilt auch für einen Großteil des sozialanthropologischen Methodenkanons: Diskursanalysen, Erzählforschung, Interviews, Literaturrecherchen, Archivquellen, Feldtagebücher – alle diese Ansätze sind vornehmlich textbasiert.[5]

Wissenschaftlich vereinheitlichte Texte sind gewiss präziser als andere kulturelle Ausdrucksformen wie etwa der Tanz, die Musik oder die Malerei. Diese Stärke ist jedoch zugleich auch eine Schwäche, denn mit Präzision lassen sich bestimmte Aspekte des Lebens schwerlich angemessen repräsentieren: Willkürliches, Chaotisches, Leidenschaftliches passt da nicht hinein und wird derart zum ausgeschlossenen ›Anderen‹.[6] Dieses Unbehagen an der Sprache als Medium und Repräsentation von Welt-Wissen und Welt-Verstehen drückte sich in den letzten Jahrzehnten in diversen Bestrebungen aus, separatistisches Denken zu überwinden: es ist hier vor allem ein Ringen mit und um die Sprache, wenn es darum geht, binäre Denkschemata zu dekonstruieren, welche unsere Welt in ein Entweder-Oder aus tot oder le-bendig, privat oder öffentlich, eigen oder fremd, Mann oder Frau, Natur oder Kultur zwingen.⁷ Man hat ständig das Gefühl, im Sprechen über diese Dichotomien, eher zu ihrer Reproduktion beizutragen statt ihnen zu entgehen. Der Vorteil des Tanzes ist es, sich ›leichtfüßig‹ über derartige Probleme hinwegsetzen zu können. Die Idee des experimentellen Tanzes liegt wiederum darin, ›diese Leichtfüßigkeit‹ für alternative Formen der Wissensproduktion ›zu nutzen‹. Hier soll dem Chaotischen, dem Zufälligen, dem Impulsiven und dem Exzessiven ein Raum gegeben werden, der nicht allein der Selbsterfahrung einzelner Teilnehmer*innen dient, sondern sich dezidiert als Beitrag zu wissenschaftlichen und ethico-politischen Diskursen versteht.⁸

---

4  *John Law/Evelyn Ruppert* (Hg.): Modes of Knowing. Ressources from the Barock. Manchester 2016, S. 29.

5  Ausnahmen bilden hier etwa ein Aufsatz von Frederike Faust und Stefan Heisenberger, in dem die Autor*innen explizit über die Rolle des Forscher*innenkörpers reflektieren: *Frederike Faust/Stefan Heisenberger:* Eine Frage des Trainings. Der Forscher_innenkörper als Erkenntnissubjekt. In: Körper-Technologien – Ethnografische und gendertheoretische Perspektiven auf die Refigurationen des Körperlichen. Berlin 2016 (= Berliner Blätter, Bd. 70), S. 68–82. Außerdem sind ethnographische Arbeiten mit audio-visuellem Material zu nennen sowie Auseinandersetzungen mit Gegenständen und Materie. Siehe etwa *Hans P. Hahn:* Vom Eigensinn der Dinge. In: Bayerisches Jahrbuch für Volkskunde 2013. München 2013, S. 13–22 oder: *Walter Leimgruber/Silke Andris/Christine Bischoff:* Visuelle Anthropologie: Bilder machen, analysieren, deuten und präsentieren. In: Sabine Hess/Johannes Moser/Maria Schwertl (Hg.): Europäisch-ethnologisches Forschen. Neue Methoden und Konzepte. Berlin 2013, S. 247–281.

6  *John Law:* After Method. Mess in Social Science Research. London/New York 2004.

Tanzen als Modus der Wissensproduktion. Das klingt für aufgeschlossene Leser*innen sicherlich erstmal interessant. Doch ist das Tanzen als solches flüchtig. Es ist lediglich in dem Moment spürbar und sichtbar, in dem es ›getan‹, also ausgeübt wird. Erst dadurch, dass über das Tanzen reflektiert, gesprochen und geschrieben wird, gewinnt es auch über den Moment hinaus an Bedeutung.⁹ Möchte man die Tanzerfahrung nicht nur als Selbsterfahrung, Sport oder ästhetischen Genuss, sondern als Modus einer weiterführenden Wissensproduktion nutzen, so muss sie in eine stabile Form gegossen werden. Aus der Erkenntnis dieser Notwendigkeit hat sich ein eigenes Genre entwickelt: das *dance-writing*.¹⁰

In *dance-writings* reflektieren Organisator*innen und Teilnehmer*innen von Movement-Research Workshops und Dance Labs zum einen über ihre sensomotorischen Erfahrungen. Dazu greifen sie mitunter auch auf literarische Mittel wie etwa die Poesie zurück. Das ist der Aspekt, der diese Texte von gewöhnlicher, wissenschaftlicher Literatur abhebt. Zum anderen werden aber auch – ganz klassisch – der Stand der Forschung, die Fragestellung, der Versuchsaufbau und das Versuchssetting dokumentiert. Darüber hinaus finden sich Verweise auf sozial- und kulturtheoretische Positionen, auf deren Grundlage weiterführende Interpretationen stattfinden. Zum Teil wird überdies im Sinne des »situated knowledge«¹¹ die spezifische, soziohistorische Perspektive reflektiert, aus welcher heraus ein *dance-writing* geschrieben wurde.

---

7   *Michael Serres/Bruno Latour:* Conversations on Science, Culture, and Time. Ann Arbor, MI 1995.
8   Webseite des *Movement Research Performace Journal*. URL: https://movementresearch.org/publications/performance-journal (Stand: 14. 2. 2020).
9   *Gabriele Klein:* Das Flüchtige. Politische Aspekte einer tanztheoretischen Figur. In: Sabine Huschka (Hg.): Wissenskultur Tanz. Historische und zeitgenössische Vermittlungsakte zwischen Praktiken und Diskursen. Bielefeld 2009, S. 199–209.
10  *Nancy Stark Smith:* Dance in Translation: The Hieroglyphs. Experiments in Writing. In: Contact Quarterly 7 (1982), Heft 2, S. 43–46.
11  *Donna Haraway:* Situated Knowledges: The Science Question in Feminism and the Privilege of Partial Perspective. In: Feminist Studies 14 (1998), Heft 3, S. 575–599.

## Über das Tanzen schreiben

Der vorliegende Text lässt sich als ein solches *dance-writing* verstehen. Hintergrund ist, dass ich selbst eine Ausbildung in zeitgenössischem Tanz abgeschlossen habe und mich am Rande einer neunmonatigen Feldforschung 2013 zu selbstorganisierten Suchttherapiegruppen in New York auch mit dem Feld des experimentellen Tanzes beschäftigt habe. Diese Beschäftigung habe ich in einer weniger intensiven Form in den letzten sechs Jahren in Deutschland fortgesetzt. Nun könnte man annehmen, dass meine Methode die ›Teilnehmende Beobachtung‹ war und diesen Text entsprechend als Autoethnographie bezeichnen.[12] Doch möchte ich bewusst einen anderen Fokus setzen. Hierzu beziehe ich mich auf das Konzept des »embodied research« nach Ben Spatz.[13]

Mit dem in den Performance Studies angesiedelten Programm der ›embodied research‹ systematisiert Ben Spatz einige der methodologischen Grundideen der Performance Art und überführt sie in eine weitergefasste Heuristik, die sich auch auf Bereiche wie zum Beispiel das Spielen eines Musikinstrumentes, das Theater oder die Bildhauerei übertragen lässt. Spatz setzt sich damit für ein Wissenschaftsverständnis ein, welches das Erlernen einer künstlerischen, sportlichen oder handwerklichen Technik als Wissensproduktion versteht und das Praktizieren dieser Technik als Forschung. Es ist hier also nicht erst das Reflektieren und Schreiben ›*über*‹ eine Technik, was als Wissenschaft gelten darf, es ist bereits die Praxis selbst. Mit Bezug auf meinen Forschungsgegenstand und -ansatz bedeutet das konkret: Ich forsche nicht nur ›*über* den Tanz*‹*. Ich forsche ›*durch das Tanzen*‹.

Tanzen ist nicht nur Forschungs›gegenstand‹. Tanzen ›ist‹ Forschung. Diese scheinbar nur leichte Verschiebung im Fokus, die das Konzept des ›embodied research‹ von der Autoethnographie abhebt, ist folgenreich. Sie rückt die Ausübung einer Praxis, das heißt die sensomotorischen Bedingungen ihrer Ausübung sowie das Training und die Fertigkeiten, derer es dafür bedarf, noch stärker in den Mittelpunkt, als es bei der Autoethnographie der Fall ist. Man könnte ›embodied research‹ so gesehen auch als einen besonders konsequenten Versuch bezeichnen, praxeologische Einsichten über ontologische Performativität in eine Methodik zu überführen.[14] Ferner begrüßt ›embodied research‹ den Prozess des ›going native‹ als integralen Bestandteil der Forschung.[15] *Going native* im Sinne des Erlernens und Praktizierens der-jenigen Technik, die das Feld konstituiert, ist im ›embodied research‹ Bedingung, Methode und reflexiver Bestandteil der Forschung, kein Fehltritt.[16]

---

12 *Brigitte Bönisch-Brednich:* Autoethnografie. Neue Ansätze zur Subjektivität in kulturanthropologischer Perspektive. In: Zeitschrift für Volkskunde 108 (2012), S. 47–63.
13 *Ben Spatz:* Embodied Research: A Methodology. In: Liminalities: A Journal of Performance Studies 13 (2017), Heft 2, o. S.
14 *Annemarie Mol:* The Body Multiple. Ontology in Medical Practice. Durham 2003.
15 *Ben Spatz:* What a Body Can Do: Technique as Knowledge, Practice as Research. New York/London 2015.

Es waren vor allem drei Typen von Movement-Research-Workshops, an denen ich teilgenommen habe und die mich in meinem Interesse für das Phänomen der Anonymität derart nachhaltig beeindruckt haben, dass ich mich dazu entschlossen habe, nach Ansätzen zu suchen, die es ermöglichen, das Tanzen selbst als Forschung zu begreifen. In allen drei Workshops spielte die Anonymität eine zentrale Rolle. Ich habe auch in anderen Kontexten[17] intensiv über Anonymität geforscht. Dabei habe ich vor allem mit klassischen, ethnographischen Methoden gearbeitet. Meine tänzerischen Erfahrungen des Einflusses, den Anonymität auf Wahrnehmungen und Gruppen haben kann, hat meine Perspektive auf den Gegenstand allerdings in einem Maße beeinflusst, welches mit keiner anderen der von mir bisher angewandten Methode vergleichbar wäre (z. B. historische und kritische Diskursanalyse, Teilnehmende Beobachtung, Netnographie, Interviews, digital metrics). Die sinnliche Erfahrung von Anonymität im Tanz prägt meine Sicht auf den Gegenstand bis heute. Im Folgenden werde ich die drei besagten Workshoptypen kurz aber dicht beschreiben, um daraufhin eine Analyse von *dance-writings* vorzustellen, bei der es darum ging, mehr über die konzeptionellen Überlegungen im Hintergrund der Workshops herauszufinden.

## Anonymität im experimentellen Tanz: drei Beispiele

### *Blind-folded Contact Dance*[18]

Der erste Workshoptyp, über den ich hier schreibe, ist zwar im Rahmen der professionalisierten Movement-Research-Bewegung entstanden, aber mittlerweile als Selbsterfahrungspraxis so populär geworden, dass es auch entgeltliche Kursangebote für Amateurtänzer*innen gibt. Der Ablauf folgt in der Regel folgendem Schema: Zunächst wärmt sich die Gruppe gemeinsam auf. Im Anschluss daran ziehen die Teilnehmenden Augenbinden an und tanzen dann in Kleingruppen einige Übungen miteinander. Während des Tanzens wissen alle Teilnehmenden trotz Augenbinden, mit wem sie gerade tanzen. Anonym im Sinne der derzeit wohl gängigsten Definition von Unerreichbarkeit[19] sind die Teilnehmenden hier demnach lediglich im optischen Sinne, da man sich gegenseitig nicht sehen kann. Darüber hinaus rücken durch die Augenbinden andere Sinneswahrnehmungen in den Vordergrund. Der meist dominante Sehsinn ist ausgeblendet. Das gegenseitige Ertasten, Erspüren, Hören und Riechen gewinnt dadurch an Relevanz. Häufig wird in diesem Zusammenhang auch mit klassischen psychoanalytischen Tropen, wie etwa der des (Ur-)Vertrauens gearbeitet.[20] Ein Beispiel hierfür sind Übungen, bei denen es um das ›Fallenlassen‹ auf der einen und das ›Auffangen‹ auf der anderen Seite geht.

---

16 Hier ist anzumerken, dass ›embodied research‹ nur auf bestimmte Felder sinnvoll anzuwenden ist.
17 Konkret bezieht sich dies auf eine ethnographische Studie zu Suchttherapiegruppen, sowie auf eine netnographische Studie zu Anonymität in Onlinechatgruppen.
18 Bei allen Workshops wird zu Beginn über einen verbindlichen Kodex zum respektvollen Umgang untereinander aufgeklärt.

### Dark-Room Contact Improvisation

Der zweite Workshoptyp geht hinsichtlich der Anonymität einen Schritt über den ersten hinaus. Hier tanzen in der Regel erfahrene Tänzer*innen ohne Unterbrechung zwei Stunden in einem komplett dunklen Raum miteinander im Stil der ›Contact Improvisation‹ (CI). Dieser lässt sich als Jazz des Tanzes bezeichnen. Es gibt hier keine vorgegebenen Choreographien. Stattdessen werden zuvor erprobte Techniken der Körper-Geist-Zentrierung, der Wahrnehmungssensibilisierung, sowie des Gewichtgebens und -nehmens angewendet. Insgesamt spielt vor allem die Behutsamkeit bei der Kontaktaufnahme eine zentrale Rolle. Bei der spezifischen Variante des ›Dark-Room-Experiments‹ agieren die Teilnehmenden dabei insofern aus der Anonymität heraus, als es im Nachhinein praktisch unmöglich ist, einschätzen zu können, wer mit wem welche Improvisationselemente getanzt hat. So ist es mir bei dieser Technik des Öfteren passiert, dass ich im Tanz eine erstaunliche Verbindung mit einem oder einer unbekannten Tanzpartner*in erlebt habe, ohne im Nachhinein herausfinden zu können, wer diese Person war, mit der ich für eine kurze Zeit eine einzigartige Form der Nähe geteilt habe.[21]

### Embodied Rebellion (becoming uncivilized)

Bei dem dritten Workshop, den ich hier diskutieren möchte, handelte es sich um eine einmalige Veranstaltung des Instituts für *Movement Research,* welches sich in New

---

19 *Helen Nissenbaum:* The Meaning of Anonymity in an Information Age. In: The Information Society 15 (1999), S. 141–144.
20 *Erik H. Erikson:* Der vollständige Lebenszyklus. Frankfurt am Main 1992.
21 Weiter gedacht könnte man diese Form der anonymen Verbindung auch als eine künstlerische Übersetzung des Konzeptes der *partial connection* interpretieren, bei der man eine Verbindung eingeht, ohne Andersartigkeit zu negieren, siehe hierzu *Marilyn Strathern:* Partial Connections. Alta Mira 2004.

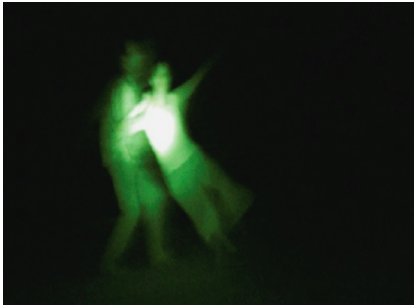 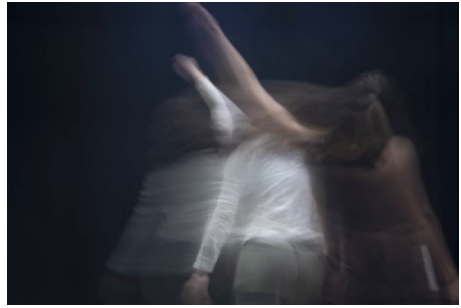

*Abb. 1 (l.): Crispin Spaeth Dance Group: Dark Room, April 8–23, 2011*
*Abb. 2 (r.): Arms to the Left. Dancers: Montana Bass and Chelo Barton. Colorado Springs, Colorado, January 2019. photo © Louisa Mackenzie*

York City befindet. Durchgeführt wurde die Veranstaltung von Luciana Achugar.[22] Achugar ist eine etablierte, uruguayische Choreographin aus der zeitgenössischen *experimental dance*-Szene. Sie nennt ihren Tanzstil *embodied rebellion*, denn es geht ihr darum, sich mit und durch den tanzenden Körper über bestimmte Schamgrenzen hinwegzusetzen, welche sie nicht nur als Resultat, sondern auch als Katalysator heteronormativer, (post-)kolonialer Unterdrückungsregime versteht. In ihrem Mission-Statement schreibt sie hierzu:

> »I make work from the rage of being a LatinAmerican living in the belly of the Empire in a post-colonial world. I make work as a practice of growing as one would grow a plant.«[23]

Hierzu arbeitet Achugar unter anderem mit dem Mittel der Anonymität. Der Workshop, an dem ich teilgenommen habe, lässt sich in vielerlei Hinsicht eher als Ritual denn als Tanzkurs verstehen, denn es geht darum, durch den Tanz einen liminalen Raum zu erzeugen, in welchem sich kulturell-kontingente Differenzen auflösen.[24] Umhüllt von schwarzen, grob-stöfflichen Planen sollten wir uns dazu in der besagten Veranstaltung durch den Proberaum bewegen. Unsere Bewegungen wurden dabei durch den groben Stoff derart verfremdet und gestört, dass wir von unseren gewohnten Bewegungsmustern abweichen mussten.

---

22  *Luciana Achugar:* About. URL: http://www.lachugar.org/about (Stand: 14. 2. 2020).
23  *Luciana Achugar:* Mission. URL: http://www.lachugar.org/mission (Stand: 14. 2. 2020).
24  Webseite des Zentrums für *Movement Research*, Ankündigung eines Workshops mit der Choreographin Luciana Achugar, URL: https://www.movementresearch.org/event/7140 (Stand: 14. 2. 2020).

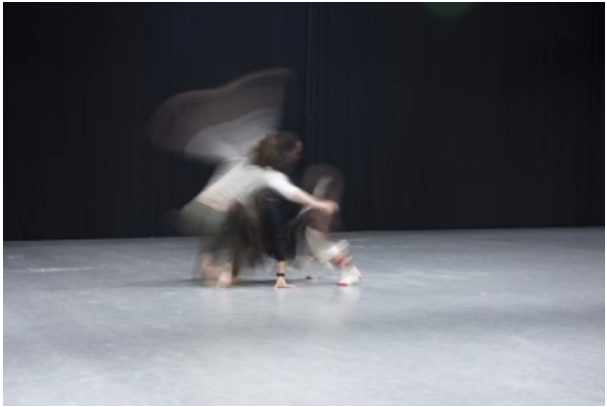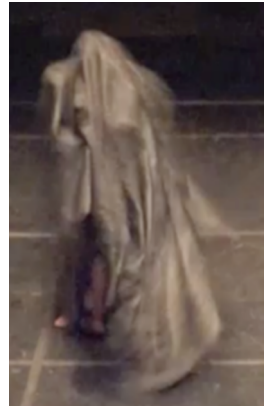

*Abb. 3 (l.): Sharing Weight. Dancers: Montana Bass and Chelo Barton. Colorado Springs, Colorado, January 2019. photo © Louisa Mackenzie*
*Abb. 4 (r.): OTRO TEATRO, All photos by © Alice Gebura, 2014*

Das Tanzen, so wie wir es in unzähligen Unterrichtsstunden erlernt hatten, war nicht mehr möglich. Wir konnten auf das Gelernte nicht zurückgreifen. Stattdessen stolperten wir, fielen hin, robbten, krochen, kauerten. Eine Teilnehmerin erlebte kurzzeitig eine panische Reaktion auf das Tuch. Sie begann zu hyperventilieren und wirr zu zappeln. Allgemein schienen die meisten sich zunächst eher unwohl zu fühlen, auch ich – bis eine gewisse Gewöhnung eintrat. Unter dem Tuch entstand im Folgenden eine besondere Form von Intimität, die eine Begegnung mit mir selbst weniger als Frau oder Mann, als Weiße oder Schwarze, als Mutter oder Tochter ermöglichte, sondern als undefiniertes Wesen – als Kreatur. Derart dekonstruiert und zugleich im Kontakt mit mir selbst, begann ich im weiteren Verlauf Verbindungen zu anderen Teilnehmer*innen aufzunehmen. Dabei sind diese Verbindungen insofern außergewöhnlich, als wir uns – verfremdet durch die Planen – nicht als Personen, sondern eher als Kreaturen adressierten. Das Workshop-Experiment ist obendrein nicht nur von Irritationen, sondern im dynamischen Wechsel auch von Freude und Komik geprägt gewesen – von Stimmungen, die wir in diversen tänzerischen ›Encounters‹ ohne Worte einander gegenüber artikulierten und miteinander teilten.

In diesen drei Experimenten wird Anonymität sehr unterschiedlich gestaltet, sie wird auf unterschiedliche Art und Weise wirksam und die Vermutung liegt nahe, dass auch unterschiedliche Überlegungen, Imaginationen und Motive hinter diesen drei Experimenten stecken. Hier kommt die Analyse von *dance-writings* ins Spiel.

*Anonymität im experimentellen Tanz: drei Auslegungsweisen*

Bei der Analyse von *dance-writings* bin ich wie folgt vorgegangen: Zunächst habe ich die letzten zehn Jahrgänge dreier Zeitschriften auf Referenzen zu Experimenten mit Anonymität durchsucht. Ich habe die Zeitschriften als Sample ausgewählt, weil sie zum einen von Schlüsselpersonen aus der Szene gegründet wurden und zum anderen, weil sie darauf spezialisiert sind, besagte *dance-writings* zu veröffentlichen und insofern als Sprachrohre der Szene gelten können. Es handelt sich erstens um ›Contact Quarterly. Journal of Dance and Improvisation‹ (CQ), zweitens um das ›Movement Research Performance Journal‹ sowie drittens um ›Entkunstung‹. Zusätzlich habe ich den Internetauftritt zweier kommerzieller Anbieter von *blind-contact dance jams* für Laien in die Analyse miteinbezogen – eine von einer deutschen Anbieterin und eine von einem kalifornischen Paar, um auch diese Form einer populärkulturellen Übersetzung bei der Analyse zu berücksichtigen. Im Folgenden diskutiere ich in Bezug auf die drei zuvor beschriebenen Workshoptypen drei unterschiedliche Auslegungsweisen der Rolle von Anonymität im Tanz. Ich beginne dafür mit der Analyse der Webpages.

*1. Tanzen als Heilung*

Die Webpages der beiden analysierten Anbieter*innen wirken professionell gestaltet. Die Professionalität wird daran ersichtlich, dass durchgängige und durchdachte Farbschemata gewählt wurden und die Seiten innerhalb eines Jahres bei regelmäßigen Aufrufen stetig aktualisiert wurden. Zusätzlich zu Informationen über die Veranstaltungen und die Veranstalter*innen sind auf den Seiten Videos integriert, die ebenso professionell geschnitten sind und einen lebendigen Eindruck von der Praxis des Tanzens selbst vermitteln. Ein dunkler Hintergrund und eine ›schummrige‹ Beleuchtung transportieren sowohl in den Videos als auch auf den Webpages insgesamt den Eindruck einer intimen, gemütlichen Atmosphäre.

Auf den Webpages der Anbieter*innen von *blind dance jams* befinden sich diverse Aussagen, die Aufschluss über mögliche Motive für und Hintergrundüberlegungen zu dem Einsatz von Augenbinden und Dunkelheit im Tanz geben können. Besonders auffällig ist die häufige Verwendung von Ausdrücken wie ›Heilung/healing‹, ›Sicher/safe‹, ›authentisch/authentic‹. Etwa schreibt eine der Organisator*innen, Candice Holdorf: »Wearing blindfolds allows dancers to drop the fear and self-judgment that often inhibits spontaneous movement and authentic connection to self

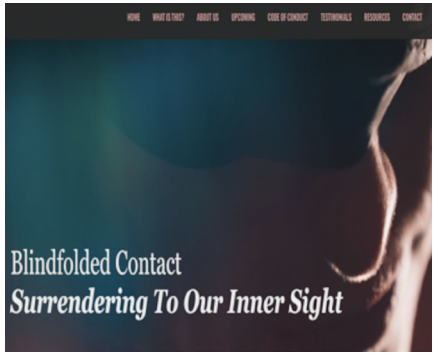 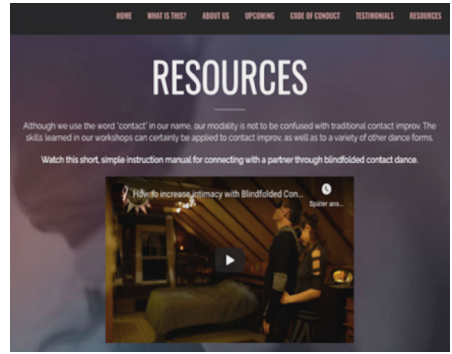

*Abb. 5/6: Screenshot: www.blindfoldedcontact.com, Stand: 1. 9. 2020, Abb. 6: Foto von: www.ganzheitlicheesundheitsförderung.de/blinddance/, Zugriff 1. 9. 2020, © Patrick Beelaert*

and others.«[25] Ferner wird auf Erfahrungsberichte von Teilnehmenden verwiesen, mit Aussagen wie: »Thanks for bringing my soul back to life. So healing.« Oder: »I came into this as a shy person that never danced. The blindfolds made me comfortable to be myself.«[26] Gabriel Diamond, ein anderer Anbieter, geht in eine ähnliche Richtung, wenn er schreibt: »Blindfolds can be used as a medicine for transcending biases, cravings, and aversions.«[27] Auf der deutschen Anbieterseite wiederum heißt es in ähnlichem Duktus:

> »Ich biete ( … ) einen Raum, um nachhause zu kommen. Um sich zu nähren, aufzuladen und aus dem Herzen zu geben. Ein Raum um sich und andere kennen und spüren zu lernen und um die Heilkraft von Berührungen im Tanz zu erfahren. Ein geschützter Raum, um loszulassen, um einfach nur zu sein, ohne Anstrengung so angenommen zu werden wie man gerade ist, sich selbst schätzen und lieben zu lernen.«[28]

Die Bezüge zu nicht näher spezifizierten Ideen von Selbst und Authentizität sowie zuvor die einer zentralisierten Seele machen deutlich, dass den hier beworbenen Veranstaltungen und den Motiven der Teilnehmenden essentialistische Vorstellungen von Selbstfindung und Seelenheilung zu Grunde liegen. Ein geteiltes Verständnis von Authentizität als bezogen auf ein vorkulturelles Selbst scheint implizit vorausgesetzt zu werden. Dieses vorkulturelle ›Selbst‹ soll im anonymen Tanz ›wiedergefunden‹ und geheilt werden. Die Rede von Heilung wiederum impliziert, dass da ›etwas Leidvolles‹ ist, was es zu lindern gilt. Was dieses Leidvolle allerdings ist und wodurch es ausgelöst wird, wird nicht näher spezifiziert. Der Anonymität wird im Zusammenhang mit der Heilung und der Rückkehr zur Authentizität eine Art Reinheitsversprechen unterstellt. Anonymität fungiert als so etwas wie eine Medizin im Rahmen einer Entgiftungskur. Bei dieser Verknüpfung von Anonymität und Reinheit handelt es sich um einen Mythos, dem man häufig in populären Diskursen zum Thema Anonymität begegnet.[29]

---

25 Webseite des Anbieters *blindfolded contact, URL*: https://www.blindfoldedcontact. com/#what-is-this (Stand: 14. 2. 2020).
26 Webseite des Anbieters *blindfolded contact.* URL: https://www.blindfoldedcontact.com/#testimonial (Stand: 14. 2. 2020).
27 *Blindfolded contact,* wie Anm. 25.
28 Webseite der Tanztherapeutin und Workshopleiterin *Steffi Rose.* URL: http://ganzheitliche gesundeitsförderung.de/blind-dance/ (Stand: 1. 9. 2020).

Im Ganzen betrachtet, das heißt hier im Zusammenspiel aus Bild, Text und Web-Design, transportieren die Internetauftritte der populären Anbieter*innen ein Bild von blind dance als einer *sicheren, heilenden* und *zentrierenden* Erfahrung. Dieses Bild steht im Kontrast zu jenem, welches bei der Analyse von dance-writings entsteht.

## 2. Tanz als Dekonstruktion

Der oben skizzierten Sichtweise stehen Reflektionen und Erfahrungsberichte gegenüber, die im Rahmen besagter *dance-writings* in den analysierten Fachzeitschriften veröffentlicht wurden. Hier wird die Erfahrung des *blind dance* in sehr viel ambivalenteren Tönen beschrieben. Zum Beispiel in folgenden Worten von Ruth Salomon, Professorin für Tanz und Theater an der University of California:

> »We experiment with blindfold dances, which invite the body to experience the anonymous politics of touching. […] There is a gulf to be crossed first. I am mute with mitten hands. […] In the swirl of dancers, it is a kaleidoscopic neurological overdrive, because my body is slovenly, tired, filled in with the dead-end exhaustion of being a singular entity anonymised on the way from here to there.«[30]

Das anonyme Tanzen erscheint hier als etwas Überforderndes, Ermüdendes, vielleicht sogar Verstörendes. Warum es sich dennoch lohnt, sich dieser Art des Tanzes zu widmen, wird bei der Lektüre weiterer *dance-writings* deutlich, die sich hierzu auf die Theorie des Dekonstruktivismus beziehen. Es wird hier – ganz im Gegenteil zu einer essentialistischen Idee von Selbstsuche und Authentizität, bei der anonymer Tanz als ein sanfter Weg zur Seelenheilung beschrieben wird – vorgeschlagen, blind CI als dekonstruktives Event zu verstehen. Dekonstruiert werden dabei zum Beispiel »pre-scribed frames of identity«.[31] Diese ›frames‹ schreiben Personen auf spezifische, hegemoniale Geschlechter- und Länderstereotypen fest. Um zu verdeutlichen, welche Rolle Tanz bei der Dekonstruktion dieser Stereotypen spielen kann, wird in einem *dance-writing* eine Textpassage von Derrida zitiert, allerdings werden dabei Derridas Überlegungen zur Praxis des Schreibens auf die Praxis des Tanzens übertragen:

> »[D]econstructive gestures appear to destabilize or cause anxiety or even hurt […] so every time I make this type of gesture, there are moments of fear. […] This doesn't happen at the moments when I'm writing [or dancing, Ergänzung d. A.]. Actually, when I write [or dance, Ergänzung d. A.]. there is a feeling of necessity, of something that is stronger than myself.«[32]

---

29 *Thorsten Thiel:* Anonymität und Demokratie. In: Paula Helm/Sandra Seubert (Hg.): Privatheit und Demokratie. Forschungsjournal Soziale Bewegungen 30 (2017), Heft 2, S. 152–161.
30 *Ruth Salomon:* Perceptions in a Contact Jam. In: Contact Quarterly 44 (2019), Heft 2.

Hier wird deutlich: Während blind dance von den einen mit einer Medizin verglichen wird, die in Form einer harmonischen Erfahrung zentrierend wirkt, beschreiben andere diesen Tanzstil als *mühselig, beängstigend* und *destabilisierend*. Allerdings wird dieser zuweilen als unangenehm empfundenen Destabilisierung eine Notwendigkeit zugerechnet, die sich aus einer gesellschaftlichen Relevanz ergibt, die wiederum darin besteht, kontingente aber durch postkoloniale Herrschaftsstrukturen zementierte Grenzen des Seins herauszufordern.

## 3. *Tanz als dekolonialisierende Praxis*

Die eine Sichtweise, das heißt anonymer Tanz als heilende Selbstfindung, und die andere, anonymer Tanz als Dekonstruktion, bilden zwei Endpunkte auf einem Kontinuum möglicher Auslegungsweisen der Rolle von Anonymität im experimentellen Tanz. Eine dritte Auslegungsweise lässt sich einem E-Mail-Interview mit Achugar entnehmen. Dieses wurde in der Zeitschrift ›Entkunstung‹ abgedruckt. Man könnte Achugars Zugang hier als einen Versuch interpretieren, die beiden vorgenannten Endpunkte konstruktiv zusammenzuführen:

---

31 *Jackie Adkins/Daniel Mang*: Still Moving: The Social Context of Contact Improvisation. A Study Lab. In: Contact Quarterly 21 (1996), Heft 2, S. 63–64.
32 *Lucía Naser* im E-Mail-Interview mit Luciana Achugar. In: Luciana Achugar: The Sublime is Us. In: Entkunstung 3 (2017). URL: https://entkunstung.com/archive/the-sublime-is-us (Stand: 12.2.2021).

> »When I let go of my ›identity‹, when I let go of being ›Luciana‹ and allow what is deep in the felt part of myself, what I came to call the ›uncivilized‹, then what emerges is more primal, animal, sexually unashamed, and powerful. […] It was coming from the desire to be lost in the experience […] and to feel empowered and unashamed of it. […] My dance is dealing with doing a healing from […] the burdens of being a Latin American living in Western civilization, ruled by colonialist thinking.«[33]

Achugar knüpft hier an die Idee einer Heilung durch das anonyme Tanzen an, begreift diesen ›Heilungsprozess‹ aber nicht als individuelle Aufgabe der Selbstfindung, sondern vielmehr als ein Aufbegehren gegen soziale Ungleichheiten, durch deren Überwindung auch persönliche ›Heilung‹ möglich wird. Damit spezifiziert sie zugleich auch das Leid, auf welches sich die Heilung durch den Tanz beziehen soll. Es ist das Leid, das Personen empfinden, die sich schämen, weil die heteronormativen Kategorien, in die sie sich fügen müssen, ihre Leidenschaften nicht respektieren. Und es ist das Leid derjenigen, die als nicht-weiß markiert sind, und als solche in bestimmte Stereotype gepresst werden, denen sie nicht entsprechen (wollen):

> »My work and ›my‹ feminism are inseparably connected/connecting being in a female-identified dancing body and being from Uruguay; my work is dealing with (practicing, doing) an embodied feminine empowered ideology from a third world country perspective that is highly aware of its postcolonial reality and its disempowered relationship to the first world. […] In the feminist theory that I›ve read, there is often a mention of how a woman›s body has a way of preventing us from being able to forget that we exist as a BODY […]. This inability and inevitability – or perhaps unwillingness and choice – to not escape the BODY in its full messy and complex nature […] this intersectional feminist proposal of the possibility of liberating ourselves from an oppressive dualist-cartesian-hierarchical way of being in the body, finds ground in the dancing body. This is why I have been calling what I do a practice of becoming uncivilized AND decolonized.«[34]

Derart sozio-historisch situiert, versteht Achugar Tanz nicht nur als Raum der heilenden Selbsterfahrung, sondern zugleich ganz dezidiert als Beitrag zu einer politischen Bewegung, als deren Ziel sie die Vervielfältigung dualistischer Geist-Körper- und Natur-Kultur-Konzeptionen sieht. In diesen Dualismen verortet sie den Ursprung heteronormativer und postkolonialer Formen der Unterdrückung, welche sie gemeinsam mit ihren Co-Tänzer*innen über den Modus der *leidenschaftlichen, chaotischen* und *lustvollen* körperlichen Erfahrung zu überwinden versucht.

---

33 *Luciana Achugar*, ebd.
34 Ebd.

### Dance with the trouble

Welche (neuen) Perspektiven eröffnet der Ansatz des embodied research? Welche Erkenntnisse über Anonymität lassen sich mit seiner Hilfe generieren? Wie fügen sich diese Erkenntnisse in bestehende wissenschaftliche Diskurse ein? Zunächst einmal lässt sich synthetisieren, dass Anonymität im experimentellen Tanz sehr unterschiedlich wahrgenommen wird. Es hängt dabei maßgeblich von den zu Grunde liegenden Überzeugungen über das Verhältnis von Selbst, Körper und Kontext ab, ob anonymes Tanzen als heilsam und zentrierend oder aber als destabilisierend und im positiven Sinne als unruhestiftend erlebt wird. In ihrer unruhestiftenden Qualität wiederum kann Anonymität im Tanz ganz dezidiert dazu eingesetzt werden, rebellische Kräfte frei zu setzen. Sie kann dazu genutzt werden, mit als einschränkend empfundenen Stereotypen und damit verbundenen Grenzen des Seins nicht nur auf individueller Basis zu ringen, sondern diese Kontingenzen kollektiv herauszufordern, auch wenn das mitunter mühevoll und schmerzhaft sein kann – dance with the trouble.[35]

Zu diesem letzten Punkt äußert Achugar, dass es ihr bei ihren Tanzexperimenten darum geht, mit und durch den Tanz utopische, lustvolle und anarchische Perspektiven auf das Sein zu entwickeln. Die Idee besteht darin, dass wir uns absichtlich in einen Zustand versetzen, in dem wir sind, ohne zu benennen, und in dem wir sehen, ohne zu wissen.[36] Aus einem künstl(er)i(s)ch erzeugten Nichtwissen heraus soll eine Öffnung gegenüber neuen Formen des Wissens entstehen. Nun könnte man meinen, dass Achugar dabei letztlich doch wieder in eine essentialistische Denkweise zurückfällt, die mittels der Anonymität zu einer vermeintlich reinen, wahren, vorkulturellen Natur des Menschseins zurück will. Aus meiner eigenen verkörperten Erfahrung heraus argumentiere ich jedoch, dass dies nicht der Fall ist. Stattdessen scheint es mir vielmehr so, dass es Achugar in der Art und Weise, wie sie Anonymität in ihrem Tanz einsetzt, um das Erweitern von Möglichkeitsräumen geht. Es geht nicht um das Erzeugen von Schutzräumen wie im populären *blind dance*, sondern um das Einreißen von Wänden.

Hierzu jedoch wird ein temporäres Überwinden eines spezifischen, historischen Modus des Menschseins notwendig, welcher mit einer Trennung von Körper und

---

35 Dies ist eine Referenz auf *Donna Haraways* Buch ›Stay with the trouble: Making Kin in the Chthulucene‹. Durham 2016, in dem Haraway dazu aufruft, sich nicht in einem lähmenden Fatalismus zu verlieren, sondern aktiv nach Möglichkeiten der Wiedergutmachung und des Widerstandes zu suchen und sich dazu mit anderen zusammenzutun.

36 *Achugar*, wie Anm. 23.

Geist ebenso wie von Natur und Kultur verbunden ist. Diese temporäre Überwindung hat weitreichende Folgen, denn auch wenn sie nur flüchtig, also für die Dauer des Tanzens erfahren wird, hinterlässt sie einen nachhaltigen Eindruck – sich durch die tänzerische Verbindung von Körper und Geist als Natur und Kultur zugleich zu ›erleben‹.[37] Für die Wissenschaftsfrage bedeutet das mit anderen Worten, dass sich eine Trennung von Natur und Kultur nicht nur im Rahmen von Studien zu ›Interspecies-Entanglements‹ oder ›Mensch-Umwelt-Relationen‹ in Frage stellen lässt. Durch den Modus des *embodied research* lassen sich ebenso gut Studien mit und über den menschlichen Körper als ›NaturenKulturen‹-Forschungen rahmen, denn hier kann deutlich gemacht werden, inwiefern wir Menschen selbst NaturenKulturen sind.[38] Die Anthropologin Natasha Myers lädt in diesem Zusammenhang beispielsweise dazu ein »to vegetalize your already more than human body«[39] und bricht damit mit der noch immer weit verbreiteten Vorstellung, dass menschliche Körper für sich genommen noch keine NaturenKulturen wären.

Myers' Aussage über die Vegetalisierung unserer immer schon mehr als menschlichen – interspecies – Körper steht im Kontext einer ethnographischen Studie über das wissenschaftliche Arbeiten mit und über Pflanzen beziehungsweise Pflanzenmolekülen. Dabei scheut sie sich nicht, auch die aus dem Munde einer Sozialanthropologin durchaus risikoreiche, ethico-politische Aufforderung »cultivate your inner-plant!« in ihre wissenschaftlichen Analysen einzuflechten. Angesichts der erstaunlichen Parallele zu Achugars Aussage »I make work as a practice of growing as one would grow a plant« überraschte es mich nicht, als ich erfuhr, dass auch Myers nicht nur Anthropologin, sondern auch Tänzerin ist. Mir scheint, um aus gewohnten Modi der Wissensproduktion und -kommunikation auszubrechen, brauchen wir mehr als nur unseren Intellekt, unser Denkvermögen, unsere Theorie. Wir brauchen auch unsere Körper. Mit all ihren Sinnen. Und ich glaube, dass es letztlich das ist, was Achugar meint, wenn sie mittels des Tanzes zu einer ›embodied rebellion‹ aufruft. Diese Rebellion richtet sich gegen Grenzen des Seins und des Wissens, die in Form kulturell-codierter Schamgefühle und Denkschemata tief in unsere Körper eingeschrieben sind, um von dort aus postkolonial durchsetzte Ungleichheiten zu konservieren und zu perpetuieren. Derartige Grenzen lassen sich wirkungsvoll nicht allein auf der Ebene der intellektuellen Repräsentation herausfordern, sondern verlangen den Einbezug ebenso lustvoller wie destabilisierender sensomotorischer Erfahrungen. Anonymität kann dabei sehr hilfreich sein, da sie eingewöhnte Muster und Verbindungen durchbricht, und derart Raum für Neues, Anderes und/oder Eigenes schafft.

---

37 *Andrea Olsen:* Body and Earth. An Experiential Guide. Middletown 2002.
38 *Frederike Gesing/Michi Knecht/Michael Flitner/Kathrin Amelang:* NaturenKulturen. Denkräume und Werkzeuge für neue politische Ökologien. Bielefeld 2018.
39 *Natasha Myers:* ›Sensing Botanical Sensoria: A Kriya for Cultivating Your Inner Plant.‹ Centre for Imaginative Ethnography (2013). URL: http://imaginativeethnography.org/imaginings/affect/sensing-botanical-sensoria (Stand: 5.2.2021).

Dr. Paula Helm
Interdisziplinäres Zentrum für Ethik in den Wissenschaften
Abteilung für Gesellschaft, Kultur und technischer Wandel (SCRATCH)
Wilhelmstr. 19
72074 Tübingen
paula.helm@uni-tuebingen.de